\documentclass[12pt]{article}
\usepackage{booktabs}
\usepackage{amsmath,amssymb,amsthm,mathrsfs,bm}
\usepackage{amscd}
\usepackage[dvipdfmx]{graphicx,color}
\usepackage{wrapfig}
\usepackage{tcolorbox}
\usepackage{float}
\usepackage{makeidx}
\usepackage[all]{xy}
\usepackage{url}
\usepackage[dvipdfmx]{hyperref}
\theoremstyle{definition}
\newtheorem{thm}{Theorem}[section]
\newtheorem*{thm*}{Theorem}

\newtheorem*{defn*}{Definition}

\newtheorem*{lem*}{Lemma}
\newtheorem{rem}[thm]{Remark}
\newtheorem*{rem*}{Remark}

\newtheorem*{con*}{Conjecture}

\newtheorem*{cor*}{Corollary}

\newtheorem*{prop*}{Proposition}

\newtheorem*{hypoth*}{Hypothesis}

\newtheorem*{claim*}{Claim}

\renewcommand{\theprf}

\newcommand*{\affaddr}[1]{#1} 

\newcommand*{\email}[1]{\texttt{#1}}
\begin{document}
\title{\bf\Large{\textsf {Hofstadter's Butterfly and Langlands Duality}\\
\noindent\rule{\textwidth}{1.5pt}}}
\author{%
\bf{\textsf Kazuki Ikeda}\\
\affaddr{\small{Department of Physics, Osaka University, Toyonaka, Osaka 560-0043, Japan}\\\small{\email{kikeda@het.phys.sci.osaka-u.ac.jp}}}
}
\date{}
\maketitle
\begin{abstract}
\hspace{-6mm}We dig out a deeper mathematical structure of the quantum Hall system from a perspective of the Langlands program. An algebraic expression of the Hamiltonian with the quantum group $\mathcal{U}_q(sl_2)$ is a cornerstone. The Langlands duality of $\mathcal{U}_q(sl_2)$ sheds light on the fractal structure of Hofstadter's butterfly. This would imply a "quantum Langlands duality". 
\end{abstract}
\newpage
\section{Introduction}
Several profound conjectures on number theory and harmonic analysis proposed by R. Langlands \cite{Langlands1970} has been developed drastically in a vast area of modern mathematics and nowadays it is recognized as the Langlands program. Those series of innovation are also directed to (quantum) integrable systems \cite{Feigin:2007mr, Frenkel:2016gxg} and representation theory of quantum groups \cite{Frenkel2011}. It has been known that the Hamiltonian \eqref{tight} of the quantum Hall effect can be written with the quantum group $\mathcal{U}_q(sl_2)$ \cite{PhysRevLett.72.1890} and its energy spectrum shows an attractive structure, called Hofstadter's butterfly \cite{PhysRevB.14.2239}, shown in figure \ref{but}. Since these seminal discoveries, it had been naively believed that there would be a more profound mathematical concept for the quantum Hall effect, whereas it had not been well understood for more than 40 years. We claim that it is nothing but the Langlands program and, in this article, we build a direct connection between Hofstadter's butterfly and the Langlands duality of $\mathcal{U}_q(sl_2)$. The integer quantum Hall effect is studied in a detailed way from a perspective of the geometric Langlands correspondence \cite{Ikeda}, which is another branch of the Langlands program.  

\paragraph*{Acknowledgement}$\\$
I thank Yasuyuki Hatsuda for providing me his Mathematica code to daw the Butterfly. I also thank Yuji Sugimoto for useful discussions. 
\section{Hofstadter's Butterfly and Lanlands Duality}
\subsection{Preliminary}
\begin{figure}[h!]
\centering
\includegraphics[width=8cm]{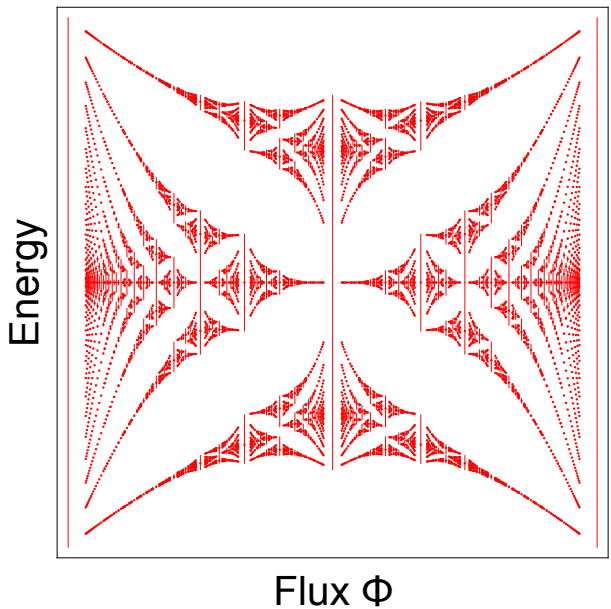}
\caption{Fractal energy spectrum, called Hofstadter's butterfly} 
\label{but}
\end{figure}

Throughout this paper, we consider the integer quantum Hall effect on a square lattice in a uniform magnetic flux $\phi=P/Q$ (in a unit of flux quantum $\phi_0=hc/e$) per plaquette, where $P$ and $Q$ are coprime numbers. Energy spectrum of the Hamiltonian \eqref{tight} shows the novel fractal structure (fig \ref{but}) as a function of the flux $\phi$, called Hofstadter's butterfly \cite{PhysRevB.14.2239}. It is known that it is generated by the maps 
\begin{align}\label{a}
\begin{aligned}
(\phi,E)&\to (\phi+1, E)\\
(\phi,E)&\to (1/\phi, f(E)),
\end{aligned}
\end{align}
where $f$ is a some function. However the origin of this fractal nature had been a mystery and hence how to determine the function $f$ had been a problem. Recently, the map \eqref{a} turned  out to have intimate relations to the modular transformation of parameters associated with a Calabi-Yau manifold \cite{Hatsuda:2016mdw, PhysRevD.95.086004}. Instead of the tight binding Hamiltonian, they worked on the Hamiltonian $H=e^{-x}+e^{x}+R^2(p^{-x}+p^{x})$ of the relativistic Toda lattice, which is a integrable system, and identified the energy spectrum as the roots of polynomials $P_\phi$. Moreover they derived the formula 
\begin{equation}\label{P}
P_\phi(E,R)=P_{1/\phi}(\widetilde{E},\widetilde{R}),
\end{equation}
where $\widetilde{E}=f(E)$ and $\widetilde{R}=R^{1/\phi}$, by using its dual Hamiltonian $\widetilde{H}=e^{-\widetilde{x}}+e^{\widetilde{x}}+\widetilde{R}^2(p^{-\widetilde{x}}+p^{\widetilde{x}})$ in terms of quantum geometry. 

\subsection{Langlands Duality of Quantum Groups}
A generic tight binding Hamiltonian we are interested in is 
\begin{equation}\label{tight}
H=\sum_{m,n}\left(c_{m+1,n}^\dagger c_{m,n}e^{A^x_{m,n}}+R^2c_{m,n+1}^\dagger c_{m,n}e^{A^y_{m,n}}+h.c.\right),
\end{equation} 
where $c_{m,n}~(c_{m,n}^\dagger)$ is the annihilation (creation) operator at $(m,n)$ site. When we choose the Landau gauge $A^x_{m,n}=0, A^y_{m,n}=m\phi$, the eigenvalues of this Hamiltonian are obtained by solving Harper's equation 
\begin{equation}\label{ha}
e^{ik_x}\psi_{m+1}+e^{-ik_x}\psi_{m-1}+2\cos(k_y+m\phi)\psi_m=E\psi_m, 
\end{equation}
where $\psi_n$ are the Bloch functions\footnote{We consider only the $x$-direction since the $y$-direction dose not contribute to the equation in the Landau gauge, which is a consequence of Bloch's theorem.} with period $Q$ $(\psi_n=\psi_{n+Q})$. Therefore our problem results in solving the characteristic polynomial associated with an equivalent  Hamiltonian which is a $Q\times Q$ matrix given by 
\begin{equation}\label{t}
H=T_x+T_x^\dagger+R^2(T_y+T^\dagger_y),
\end{equation}
where we choose a $Q$-dimensional representation $\rho_Q$ of $\mathcal{U}_q(sl_2)=\{K^{\pm1},X^\pm\}$ with $q=e^{i\pi P/Q}$ so that
\begin{align}\label{sl}
\begin{aligned}
T_x&=e^{ik_x}\rho_Q(X^+),~~T_y=e^{ik_y}\rho_Q(K)\\
\rho_Q(X^+)&=\begin{pmatrix}
0&1&0&\cdots&0\\
\vdots&\ddots&\ddots&\ddots\\
\vdots&&\ddots&\ddots&0\\
0&&&\ddots&1\\
1&0&\cdots&\cdots&0
\end{pmatrix}
,~~\rho_Q(K)=\text{diag}(q^2,q^4,\cdots,q^{2Q})
\end{aligned}
\end{align}
These operators $T_x$ and $T_y$ are non commutative because of the Aharonov-Bohm phase for an electron moving around the flux:
\begin{equation}
T_xT_y=q^2T_yT_x. 
\end{equation} 
The energy spectrum consists of eigenvalues of this Hamiltonian, which is described by the Chambers relation \cite{PhysRev.140.A135}
\begin{equation}
\det(H(k,R)-E)=P_\phi(E,R)+h(k,R),~~k=(k_x,k_y)
\end{equation}
where $P_\phi(E,R)$ is a polynomial and $h(k,R)=2(-1)^{Q-1}(\cos(Qk_x)+R^{2Q}\cos(Qk_y))$. The point $k_0=(\pi/2Q,\pi/2Q)$ where $h(k,R)$ vanishes is called a mid band point. Therefore the energy spectrum at the mid band point oveys $P_\phi(E,R)=0$. Some examples of $P_\phi$ are as follows: 
\begin{align}
\begin{aligned}
P_{P/1}(E,R)&=E\\
P_{1/2}(E,R)&=E^2-2(1+R^4)\\
P_{1/3}(E,R)&=P_{2/3}(E,R)=E(-E^2+3+3R^4)\\
P_{1/4}(E,R)&=P_{3/4}(E,R)=E^4- 4 (1 + R^4)E^2+2 (1 + R^8)  \\
P_{1/5}(E,R)&=P_{4/5}(E,R)=- E^5+ 5 (1 + R^4) E^3 +\frac{5}{2}(-2 + (-3 + \sqrt{5}) R^4 - 2R^8)E\\
P_{2/5}(E,R)&=P_{3/5}(E,R)=- E^5+ 5 (1 + R^4) E^3 -\frac{5}{2} (2 + (3 + \sqrt{5}) R^4 + 2 R^8) E
\end{aligned}
\end{align}

The above polynomials are exactly the same as in \cite{Hatsuda:2016mdw}. Hence the anticipated formula $P_{P/Q}(E,R)=P_{Q/P}(\widetilde{E},\widetilde{R})$ \eqref{P} implies the equivalence of the $Q$-dimensional representation \eqref{sl} of $\mathcal{U}_q(sl_2)$ and the $P$-dimensional representation of $\mathcal{U}_{^Lq}(sl_2)$, where $^Lq=e^{i\pi/\phi}$ and $\mathcal{U}_{^Lq}(sl_2)$ is the Langlands dual quantum group of $\mathcal{U}_q(sl_2)$ \cite{Faddeev:1999fe, Frenkel2011}. We write this duality map by 
\begin{equation}\label{z}
\widetilde{S}:(\mathcal{U}_q(sl_2), H)\to (\mathcal{U}_{^Lq}(sl_2), \widetilde{H}),
\end{equation}
where the dual Hamiltonian $\widetilde{H}$ is given by the following $P\times P$ matrix of the form 
\begin{equation}
\widetilde{H}=\widetilde{T}_x+\widetilde{T}_x^\dagger+\widetilde{R}^2(\widetilde{T}_y+\widetilde{T}^\dagger_y),
\end{equation}
where $\widetilde{T}_x=e^{i\widetilde{k}_x}\rho_P(X)$ and $\widetilde{T}_y=e^{i\widetilde{k}_y}\rho_P(Y)$. Since we expect the correspondence of the characteristic polynomials $\det(H-E)=\det(\widetilde{H}-\widetilde{E})$, we find $\widetilde{R}=R^{1/\phi}$ by comparing order of $R$ and $\widetilde{R}$ in $h(k,R)$ and $\widetilde{h}(\widetilde{k},\widetilde{R})$.

The reason why this duality originates from the Langlands duality\footnote{This viewpoint would differ from the modular doublet of quantum groups \cite{Faddeev:1999fe}.} of quantum groups is explained by the interpolating quantum group $\mathcal{U}_{q,t}(sl_2)$ \cite{Frenkel2011}, which is parametrized by arbitrary nonzero complex values $q,t$ and generated by $X^\pm,K^{\pm1},\widetilde{K}^{\pm{1}}$ such that
\begin{align}
\begin{aligned}
KX^\pm&=q^{\pm2}X^\pm K,~~\widetilde{K}X^\pm=t^{\pm2 }X^\pm \widetilde{K},\\
[X^+,X^-]&=\frac{K\widetilde{K}-(K\widetilde{K})^{-1}}{qt-(qt)^{-1}}.
\end{aligned}
\end{align}
The interpolating property of $\mathcal{U}_{q,t}(sl_2)$ appears as 
\begin{equation}
\mathcal{U}_{q,1}(sl_2)/\{\widetilde{K}=1\}\simeq \mathcal{U}_q(sl_2),~~\mathcal{U}_{1,t}(sl_2)/\{K=1\}\simeq \mathcal{U}_t(sl_2). 
\end{equation}
By definition, $\mathcal{U}_{q,t}(sl_2)$ is equivalent to the usual quantum group $\mathcal{U}_\varrho(sl_2)$ with generators $X^{\pm}, K\widetilde{K}$ and the parameter $\varrho=qt$. Taking $q=e^{i\pi P/Q}$ and $t=\hspace{-2mm}~^Lq=e^{i\pi Q/P}$, we find $\varrho=q~^Lq=e^{i\pi(P/Q+Q/P)}$ is symmetric under exchanging $P$ and $Q$. Therefore the fractal structure in Hofstadter's butterfly embodies "symmetry breaking" of this quantum group $\mathcal{U}_\varrho(sl_2)$ into a quantum group $\mathcal{U}_q(sl_2)$ and its Langlands dual quantum group $\mathcal{U}_{^Lq}(sl_2)$. The Langlands duality of quantum groups is formulated by E. Frenkel and D. Hernandez \cite{Frenkel2011} and, according to which, any irreducible representation of $\mathcal{U}_{q}(sl_2)$ would be $t$-deformed uniquely to a representation of $\mathcal{U}_{q,t}(sl_2)$ in such a way that its specialization at $q=1$ gives a representation of $\mathcal{U}_{t}(sl_2)$. The easiest case is $P=1$ and $Q=2$. A two-dimensional representation of $\mathcal{U}_q(sl_2)$ is dual to a one-dimensional representation of $\mathcal{U}_{t}(sl_2)~(t=\hspace{-1mm}^Lq=e^{\pi i Q/P})$, which is equivalent to $P_{1/2}(E,R)=P_{2/1}(\widetilde{E},\widetilde{R})$. Generically, we observe that a $Q$-dimensional representation of $\mathcal{U}_{q}(sl_2)$ and a $P$-dimensional representation of $\mathcal{U}_{^Lq}(sl_2)$ are dual. 

\subsection{Discussion}
Let us conclude this article with comments on recent advances of the Langlands duality.
The quantum affine algebra $\mathcal{U}_q(\widehat{sl_2})$ is natural extension of $\mathcal{U}_q(sl_2)$ and the deformed $\mathcal{W}$-algebra $\mathcal{W}_{q,t}(sl_2)$ interpolates $\mathcal{U}_q(\widehat{sl_2})$ and $\mathcal{U}_t(\widehat{sl_2})$ \cite{1998math.....10055F}, where $t=q^\beta$ with a parameter $\beta$. More recently the Langlands duality between a conformal block of $\mathcal{U}_q(\widehat{sl_2})$ and that of $\mathcal{W}_{q,t}(sl_2)$ was established in \cite{ Aganagic:2017smx}. In the limit of $q\to1$ with $t=q^\beta$, we have $\mathcal{W}_{q,t}(sl_2)\to \mathcal{W}_{\beta}(sl_2)$ \cite{1997q.alg.....8006F}. What is called the Langlands duality of $\mathcal{W}$-algebras \cite{FF} is 
\begin{equation}
\mathcal{W}_\beta(sl_2)=\mathcal{W}_{^L\beta}(sl_2),~~\beta ^L\beta=1.
\end{equation}
Surprisingly, they are related to $S$-duality of four dimensional supersymmetric Yang-Mills theories with parameters $\beta=R_s/R_t$ and $^L\beta=R_t/R_s$, where $R_s$ and $R_t$ are radii of circles in $T^2=S^1_s\times S^1_t$  \cite{Aganagic:2017smx}. In our case these parameters are somehow identified with $R_s=P$ and $R_t=Q$. Taking $q=e^{i\pi\beta}$ and $t=\hspace{-1mm}^Lq=e^{i\pi ^L\beta}$, we have the following duality picture:
\begin{equation}
\mathcal{U}_{t}\left(\widehat{sl_2}\right)\leftrightarrow \mathcal{W}_{q,t}(sl_2)\to\mathcal{W}_{\beta}(sl_2)=\mathcal{W}_{^L\beta}(sl_2)\gets\mathcal{W}_{t,q}(sl_2)\leftrightarrow\mathcal{U}_{q}\left(\widehat{sl_2}\right).
\end{equation}
These relations with supper Yang-Mills theories are quite interesting, since it implies some relations with the geometric Langlands duality in an unexpected way. It desires further endeavor to clarify all of those relations.    

\begin{rem}
We leave comments on relation to $S$-duality of four-dimensional $\mathcal{N}=4$ suppersymmetric Yang-Mills theories. In physics, the geometric Langlands correspondence was firstly mentioned for mirror symmetries of Hitchin moduli spaces $\mathcal{M}_G$ and $\mathcal{M}_{^LG}$ as results of dimensional reduction to a two-dimensional sigma model \cite{Kapustin:2006pk}. These Hitchin moduli spaces are exchanged by $S$-transformation of the form
\begin{equation}
S:(G,\tau)\to(^LG,1/n_\mathfrak{g}\tau),
\end{equation}
where $^LG$ is the Langlands dual Lie group\footnote{P. Goddard, J. Nuyts, and D. Olive found the Langlands dual groups from physical motivation \cite{GODDARD19771}. $^LG$ is known as a GNO dual group among physicists. } of $G$, $\tau$ is a parameter of the theory, and $n_\mathfrak{g}$ is the ratio of length squared of long and short roots of $G$. What is different in our case is the $\widetilde{S}$-transformation \eqref{z} plays an alternative role. Hence we interpret that $\widetilde{S}$-duality endows another mathematical phenomenon, which we may call a "quantum Langlands correspondence".     
\end{rem}
\if{
\subsection{Bethe Ansatz and the Energy Spectrum}
Now let us determine the shape of the unknown function $f$ in \eqref{a}. For this purpose it is paramount of importance to recall the relations between Hofstader's butterfly and the Bethe ansatz equations \cite{PhysRevLett.72.1890, 1996PhRvB..53.9697H, 2000PhRvB..61.4409H}. We consider the case $\phi=P/Q$ and then there are $Q$ energy bands. By choosing a certain gauge, Harper's equation \eqref{ha} can be equivalently deformed to the equation 
\begin{equation}
i(z^{-1}+qz)\Psi(qz)-i(z^{-1}+q^{-1}z)\Psi(q^{-1}z)=E\Psi(z),
\end{equation} 
where $z$ is a parameter such that $\psi_n=\Psi(q^n)$, which are periodic in 2Q $(\psi_n=\psi_{n+2Q})$. This $\Psi(z)$ is a polynomial of degree $Q-1$ and the set of zero points $\{z_i\}$ of $\Psi(z)$ obey the Bethe ansatz equation
\begin{equation}
\frac{z^2_n+q}{qz^2_n}=-\prod_{m=1}^{Q-1}\frac{qz_n-z_m}{z_n-qz_m},~~n=1,\cdots,Q-1.
\end{equation}
Moreover the spectrum is given by the sum of roots
\begin{equation}
E=-i(q-q^{-1})\sum_{n=1}^{Q-1}z_l. 
\end{equation}
Since the above derivation does not depend on the numerator of $\phi=P/Q$, we just exchange replace $Q$ by $P$ to obtain the dual energy $\widetilde{E}$. Hence $\widetilde{E}$ can be written by the sum of the roots of the "dual" Bethe ansatz equation. In principle we can determine $f$ by this procedure.  
}\fi

\bibliographystyle{utphys}
\bibliography{Ref}
\end{document}